August 9, 2007

# Electron Thermal Microscopy

Todd Brintlinger[1,2], Yi Qi[1], Kamal H. Baloch[3], David Goldhaber-Gordon[4], and John Cumings[1,4*]

[1]Department of Materials Science and Engineering, [2]Center for Superconductivity Research, and [3]Institute for Physical Sciences and Technology, University of Maryland, College Park, MD 20742
[4]Department of Physics, Stanford University, Stanford, CA 94305
*Corresponding author: cumings@umd.edu

**The progress of semiconductor electronics toward ever-smaller length scales and associated higher power densities brings a need for new high-resolution thermal microscopy techniques[1,2]. Traditional thermal microscopy is performed by detecting infrared radiation with far-field optics, where the resolution is limited by the wavelength of the light[3]. By adopting a serial, local-probe approach, near-field and scanned-probe microscopies can surpass this limit but sacrifice imaging speed[4-15]. In the same way that electron microscopy was invented to overcome the resolution limits of light microscopy[16], we here demonstrate a thermal imaging technique that uses an electron microscope to overcome the limits of infrared thermal microscopy, without compromising imaging speed. With this new technique, which we call electron thermal microscopy, temperature is resolved by detecting the liquid-solid transition of arrays of nanoscale islands, producing thermal maps in real-time (30 thermal images per second over a 16μm$^2$ field-of-view). The experimental demonstration is supported by combined electrical and thermal modeling.**

A key challenge in pushing thermal imaging to smaller length scales is generating reliable information about the local temperature. For instance, scanning thermal microscopy[7] infers the sample temperature from interfacial heat transfer, which then must be separately characterized due to its strong dependence on experimental details (e.g. thermal conductivities, probe shape, water meniscus etc.). Near-field optical techniques[8] similarly must rely upon a given surface evanescent emissivity, or characterize it separately. To provide a direct measure of temperature, we demonstrate that nanoscale metal islands can be used as passive local temperature probes. Observing phase transitions in an ensemble of islands gives a map of the absolute temperature relative to the transition-point. These transitions can be standardized[17], and provide a reliable measure of the local thermodynamic temperature, by virtue of being connected to intrinsic equilibrium phase-phenomena. Furthermore, each island is sampling temperature simultaneously, allowing a parallel determination of temperature over the entire field-of-view. This is in contrast to serial techniques which require that the probe come to equilibrium at each point separately, making electron thermal microscopy inherently faster than any serial technique. This measurement platform is applicable to a variety of materials, is complementary to existing measurement techniques, and has potential for evaluating thermal management as devices scale down to lengths approaching characteristic scattering distances of thermal carriers, typically[18] 10-100nm.

Figure 1 illustrates the principle underlying electron thermal microscopy[19]. Fig. 1a depicts islands of a low-melting-point metal (here, indium) deposited on the back side of a 100nm thick, commercially available silicon nitride membrane[20]. Using a specimen heating holder, the metal islands are heated through the melting transition in-situ while imaging in a transmission electron microscope. Interestingly, the island shape does not change upon heating through the melting transition, owing to an oxide crust encasing each island. There is also no detectable change in contrast during melting when the islands are imaged in standard bright-field mode. However, the composite diffraction pattern shown in Fig. 1b for both liquid and solid phases reveals a marked change in the diffraction properties of the islands between the two. Tilting the incident imaging beam so that the objective aperture is at the position indicated by the white circle in Fig. 1b yields the dark-field images shown in (c) and (d) for the solid and liquid phases, respectively. In this imaging mode, the melting transition of each island is easily detectable as a change in the intensity at the location of the island. Therefore, it serves as a binary thermal imaging technique, with each metal dot reporting the local temperature as either above or below $T_{melt}$. The melting temperatures of the individual dots are all close to the bulk melting temperature of indium (156.6° C), with a small amount of melting-point suppression (typically less than 10° C), depending on the size and shape of the islands[21]. This suppression is more significant in smaller diameter islands, and is limited in the present study by using only larger islands (>30nm in-plane radius) for temperature determination. Upon cooling, the dots solidify once again at or below their initial melting temperature (see Supplementary Discussion). These phenomena are in agreement with previous reports[22,23]. The metal is thus robust to thermal cycling, and allows repeated temperature measurements of a given region.

Imaging this transition in many islands over a wide field-of-view, it is possible to study local temperature changes on the non-etched front side of the silicon nitride membrane. To create local temperature gradients, metal heater wires were fabricated on the front side of the membranes and biased to create Joule heating. A bright-field image of the device is shown in Fig. 2a with tapered electrodes at top and bottom and a 54 nm wide, 870 nm long heater line spanning between them. Fig. 2b is a thermal map derived from a series of thermal images, such as those in Figs. 1c and 1d, applied to the region in Fig 2a. The color of each island is determined by the lowest heater current that causes it to melt; thus, areas of like color in the thermal image represent isothermal regions of the specimen. The hottest regions occur in the middle of the heater wire, as expected for Joule heating in standard Ohmic metal wires.

Figure 2c depicts modeling of the device using finite-element analysis[24] with colors here taking on the same meaning as in Fig. 2b. In this model, described further in the Methods section, temperature is given by:

$$\nabla \cdot (K \nabla T) + Power = 0 \qquad (1)$$

where *K* is the local thermal conductivity, *T* is the temperature, and *Power* is given by,

$$Power = \sigma |\nabla V|^2. \qquad (2)$$

Here, *V* is the electric potential, and σ is the electrical conductivity. The temperature dependence of σ is given by:

$$\sigma = \sigma_0 (1 + \alpha \Delta T)^{-1}, \qquad (3)$$

where $\sigma_0$ is the room temperature value and α is the temperature coefficient of resistivity. Potential and conductivity are governed by the steady-state equation:

$$\nabla \cdot (\sigma \cdot \nabla V) = 0. \qquad (4)$$

To test the model we leave the thermal conductivity of the silicon nitride as a free parameter. The lowest voltage needed to initiate melting matches the experimental value when we set the modeled thermal conductivity of the silicon nitride to be $3.6_{-0.1}^{+0.5}$ W/m·K. (The asymmetric uncertainty stems from melting point suppression, as explained in the Supplementary Discussion.) This value is then used to produce the thermal map shown in Fig. 2c. Our inferred thermal conductivity agrees with previous measurements on free-standing amorphous silicon nitride membranes[25-28].

The influence of the probe on the measurement is a potential concern for any thermal microscopy. For our new technique, the heating effect of the electron beam is easily quantified experimentally. Turning on the beam current for imaging typically adds 3.0 pA of current to the heater circuit. By reducing the temperature of the electron source, we can reduce the beam current by more than a factor of 30, so that this additional current falls below the noise level of our instrumentation (0.1 pA). At this reduced beam intensity, dark-field imaging is still possible, and the ensemble of indium islands systematically requires $1.0 \pm 0.6$ μW more Joule-heating power to melt than under conventional imaging conditions. This small amount of power is consistent with the amount that would be generated by 3.0 pA of current from a 300 kV imaging beam, suggesting that a majority of the beam-induced heating stems from high-momentum, high-loss scattering of the imaging electrons that eventually manifests itself as an electrical current in the electrodes. The beam-induced heating corresponds to a maximum of a few percent of the overall heating power used here, showing that our 'probe' (imaging electron beam together with nanoscale metal islands) is quantifiably non-invasive.

In our demonstration we also measure the time-resolution. Figure 3 shows the time response of an ensemble of approximately 50 islands averaged over 14 repetitions of a step-function melting transition as imaged using video acquisition. At time t = 0, a bias large enough to melt all islands in the field of view is applied to a heater wire such as that in Fig. 2. The intensity of each island is then measured in each video frame. The y-axis of Fig. 3 gives the average intensity as measured over all islands in the field of view, with 0 and 1 representing the intensity of the solid and liquid phases, respectively. The step size along the x-axis corresponds to the 60 Hz acquisition rate in the video capture system. A fit yields an exponential time constant near 30ms for a typical island. A thermal model based on a thermal diffusivity of silicon nitride of 0.025-0.25 $cm^2/s$ and a field-of-view of several microns suggests that the thermal response time of our sample to a step function of input power should be from 1-100 μsec. Therefore, thermal diffusion is not rate-determining in our thermal imaging, and the measured time response is probably limited by the image acquisition system. As alluded to previously, this time response is for a given island within the entire field-of-view. Separate islands respond this quickly, but at spatially-distinct points within a frame.

The spatial resolution of the technique is determined by the spacing of the metal islands and by the thickness and diffusivity of the dielectric membrane that separates them from the heat source to be imaged, here 100 nm. While island diameters in this study ranged from <100 nm up to 200 nm, it is possible to achieve smaller, more spherical, and more uniform diameter islands by use of different substrate temperatures during metal deposition[29]. At smaller sizes, melting point suppression of small particles implies a tradeoff between spatial resolution and temperature resolution[21,29], as uncertainties in diameter produce uncertainties in melting temperature. However, achieving a spatial resolution of 30 nm together with a temperature resolution of 10° C is reasonably achievable. Silicon nitride membranes in this thickness range are also commercially available. We also point out that the metal islands might instead be synthesized by chemical methods with even smaller sizes and size distributions[30]. In this way, the spatial resolution could be enhanced without sacrificing temperature resolution.

In summary, we present an electron imaging technique that produces thermal maps with up to 16μm$^2$ fields-of-view, video-rate acquisition, and a spatial resolution firmly in the nanoscale range. We expect the technique to be useful as a method for observing heat generation, dissipation, and transport in a variety of nanoscale systems. While the demonstration here utilizes a transmission electron microscope, which is only applicable to electron-transparent device geometries, preliminary results indicate that the technique can be readily adapted to scanning electron microscopy and back-scattered electron detection. Such thermal microscopy could be readily incorporated into a routine device-characterization protocol for a broad range of technologically-relevant geometries.

**Methods:**

Heater wires are fabricated using conventional electron beam lithography on 100 nm thick, 500 x 500 μm$^2$ silicon nitride membrane windows inside ~3 x 3 mm$^2$, 500 μm thick silicon substrates (purchased from SPI Supplies). For the lithography, membranes are spin-coated at 6 krpm with a thin layer of 950,000 molecular weight PMMA resist dissolved to 2% by weight in chlorobenzene (Microchem, 950KPMMAC2), followed by electron-beam exposure in a Philips XL30 scanning electron microscope fitted with an NPGS lithography system. The electron dose for exposure was 250 μC/cm$^2$, and focusing was aided by a layer of conducting polymer, AquaSAVE from Mitsubishi Rayon. After developing the PMMA resist for 60 sec in methyl-isobutyl ketone diluted 3:1 by volume with isopropyl alcohol, pattern transfer is accomplished by deposition of electrode material (3-5 nm titanium, 26-30 nm palladium) in an electron-beam evaporator, followed by acetone lift-off. Indium (99.997% purity) is then deposited on the back side of the membrane by thermal evaporation in ~1 x 10$^{-6}$ torr, with the substrate held at room temperature and 18nm thickness given by adjacent crystal monitor. A transmission electron micrograph of a typical final device is shown in Fig. 2a.

Experiments are performed in either a J2100 or a J4000FX JEOL transmission electron microscope operating at 200 and 300kV, respectively. Each utilizes a custom-built in-situ electrical-measurement specimen holder together with a Keithley 236 source-measure unit. Images are acquired with either a Gatan Orius SC1000 CCD or a Gatan 622-0600 for the J2100 or J4000FX systems, respectively. To compile the map shown in Fig. 2b, a series of images are recorded for 1mV increments in applied bias (approximately 1.8µA). At each increment, a low-bias, solid-island image is also acquired. These images are numerically subtracted to determine which islands melt at each bias. The pixels corresponding to the molten islands are assigned the value of the current at which they melt, while the remaining pixels are assigned the value of the island nearest to them. This results in the map seen in Fig. 2b, where current values correspond to isothermal regions for a given current.

Simulations are performed in the FlexPDE v4.2 finite-element model simulation environment. The geometry is the same as the actual device, and calculation takes place over the entire 500 x 500 µm$^2$ membrane window, with the edge of the window acting as the thermal bath. From this geometry, a mesh of 7345 nodes and 3620 cells is built. The steady-state solutions to equations (1), (2), and (4) are then computed iteratively, using equation (3) to determine electrode thermal conductivity with α determined as described in Supplementary Discussion, and the thermal conductivity of the nitride matched to experiment as described in the text. The iteration end-point is defined by a FlexPDE ERR parameter of 0.0003. Plots of temperature distribution are generated for a given set of biases, and the 157 °C regions ($T_{melt}$ of indium) of these temperature plots are then combined to create Fig. 2c.

**Acknowledgements:**

The authors thank S. J. Chen for discussions about dark-field imaging, A. Marshall, H. Furukawa and M. Kawasaki for technical assistance, O. Mills for specimen heating holder and J. Carpenter for preliminary electron back-scattering diffraction work. This work was supported by U.S. Air Force Grants No. FA9550-04-1-0384 and No. F49620-02-1-0383, and ONR Grant No. N00014-02-1-0986. Preliminary devices were fabricated at the Stanford Nanofabrication Facility of NNIN, supported by the National Science Foundation under Grant No. ECS-9731293. Measurements on those devices were completed at Stanford. All subsequent work was done at the Maryland Nanocenter, and supported by the University of Maryland.

**Author contributions:**

T.B. designed and fabricated samples, performed thermal microscopy, compiled data and provided analysis. Y.Q. wrote and ran simulation and modeling. K.H.B. designed and assembled *in situ* electrical measurement holder. D.G.G. aided in experiment conception and preliminary data acquisition. J.C. conceived and performed experiments, as well as data compilation and analysis. T.B. and J.C. co-wrote the paper. All authors commented on manuscript and aided in data interpretation.

**Competing financial interest**

Authors declare no competing financial interest.

**References:**

1. International Technology Roadmap for Semiconductors. *http://public.itrs.net* (2005).
2. Cahill, D. G. *et al*. Nanoscale thermal transport. *J Appl Phys* **93**, 793 (2003).
3. Incropera, F. P. & DeWitt, D. P. *Fundamentals of heat and mass transfer*, 5th ed. (J. Wiley, New York, 2002).
4. Nakayama, Y. *et al*. Tunable nanowire nonlinear optical probe. *Nature* **447**, 1098 (2007).
5. Thomas, J. Optics: Nanowire lasers get in tune. *Nat Nano* **2**, 392 (2007).
6. De Wilde, Y. *et al*. Thermal radiation scanning tunnelling microscopy. *Nature* **444**, 740 (2006).
7. Majumdar, A. Scanning thermal microscopy. *Annu Rev Mater Sci* **29**, 505 (1999).
8. Betzig, E. & Trautman, J. K. Near-field optics: Microscopy, spectroscopy, and surface modification beyond the diffraction limit. *Science* **257**, 189 (1992).
9. Shi, L. & Majumdar, A. Recent developments in micro and nanoscale thermometry. *Microscale Therm Eng* **5**, 251 (2001).
10. Hammiche, A., Reading, M., Pollock, H. M., Song, M., & Hourston, D. J. Localized thermal analysis using a miniaturized resistive probe. *Rev Sci Instrum* **67**, 4268 (1996).
11. Edinger, K., Gotszalk, T., & Rangelow, I. W. Novel high resolution scanning thermal probe. *J Vac Sci Technol B* **19**, 2856 (2001).
12. Nonnenmacher, M. & Wickramasinghe, H. K. Scanning probe microscopy of thermal-conductivity and subsurface properties. *Appl Phys Lett* **61**, 168 (1992).
13. Olson, E. A. *et al*. Scanning calorimeter for nanoliter-scale liquid samples. *Appl Phys Lett* **77**, 2671 (2000).
14. Kuball, M. *et al*. Measurement of temperature in active high-power AlGaN/GaN HFETs using Raman spectroscopy. *Electron Device Letters, IEEE* **23**, 7 (2002).
15. Grabiec, P. *et al*. Batch fabricated scanning near field optical microscope/atomic force microscopy microprobe integrated with piezoresistive cantilever beam with highly reproducible focused ion beam micromachined aperture. *J Vac Sci Technol B* **22**, 16 (2004).
16. Ruska, E**.** The development of the electron-microscope and of electron-microscopy. *Rev Mod Phys* **59,** 627 (1987).
17. Preston-Thomas, H. The international temperature scale of 1990 (ITS-90). *Metrologia* **27**, 3 (1990).
18. Cahill, D. G., Goodson, K., & Majumdar, A. Thermometry and thermal transport in micro/nanoscale solid-state devices and structures. *J Heat Trans* **124**, 223 (2002).
19. Lereah, Y., Deutscher, G., & Kofman, R. Evidence for pre-melting surface loosening of lead clusters embedded in a silicon monoxide matrix. *Europhys Lett* **8**, 53 (1989).
20. Membranes obtained from SPI company, available at http://www.2spi.com/.
21. Allen, G. L., Bayles, R. A., Gile, W. W., & Jesser, W. A. Small particle melting of pure metals. *Thin Solid Films* **144**, 297 (1986).
22. Blackman, M., Sambles, J. R., & Peppiatt, S. J. Superheating of bismuth. *Nature-Physical Science* **239**, 61 (1972).

23. Olson, E. A., Efremov, M. Y., Zhang, M., Zhang, Z., & Allen, L. H. Size-dependent melting of Bi nanoparticles. *J Appl Phys* **97** (2005).
24. FlexPDE® version 4.2 (PDESolutions, Inc., 2006).
25. Mastrangelo, C. H., Tai, Y. C., & Muller, R. S. Thermophysical properties of low-residual stress, silicon-rich, LPCVD silicon-nitride films. *Sensor Actuat a-Phys* **23**, 856 (1990).
26. Lee, S. M. & Cahill, D. G. Heat transport in thin dielectric films. *J Appl Phys* **81**, 2590 (1997).
27. Holmes, W., Gildemeister, J. M., Richards, P. L., & Kotsubo, V. Measurements of thermal transport in low stress silicon nitride films. *Appl Phys Lett* **72**, 2250 (1998).
28. Zink, B. L. & Hellman, F. Specific heat and thermal conductivity of low-stress amorphous Si-N membranes. *Solid State Commun* **129**, 199 (2004).
29. Wronski, C. R. M. Size dependence of melting point of small particles of tin. *British Journal of Applied Physics* **18**, 1731 (1967).
30. Schmid, G. Large clusters and colloids. Metals in the embryonic state. *Chemical Reviews* **92**, 1709 (1992).

**Figure legends:**

**Figure 1.** Imaging liquid/solid phase transitions in a transmission electron microscope. a.) A bright-field image of indium islands deposited on a silicon nitride membrane. b.) A composite diffraction pattern from indium islands in both liquid and solid phases. The white circle shows approximately where the objective aperture is placed for the dark-field images in c) and d). c) and d) Dark-field images of same region with indium islands in the solid and liquid phases, respectively. Thus, each island acts as a local, binary temperature probe. Scale bar is 300nm.

**Figure 2.** Demonstration of electron thermal microscopy. a) A bright-field transmission electron micrograph of a heater wire with tapered electrodes fabricated on a silicon nitride membrane using electron-beam lithography. Application of bias to the electrodes allows nanoscale thermal gradients to be produced near the wire due to Joule heating. Indium islands are visible on the back side of the membrane. Scale bar is 1 µm. b) A thermal map of the same region. Each pixel is colored according to the bias current needed to melt the indium island nearest to that pixel. The map is assembled from 50 separate images, recorded at increments of applied bias. c) A finite-element thermal model of the device using a thermal conductivity of 3.6 W/m·K for the silicon nitride and a temperature coefficient of resistivity, α, of $1.8 \times 10^{-3}$/K for palladium. Here, the colors represent currents that cause the regions to be greater than 157°C, the melting temperature of bulk indium.

**Figure 3.** Time response of individual indium islands to a rapid temperature change. Y-axis indicates value assigned to contrast within a given indium island, with 0 and 1 representing solid and liquid, respectively. X-axis is time, with t = 0 representing application of a voltage sufficient to melt all islands in field-of-view. Error bars show standard deviation over 14 different sets of images. Fitting (gray line) to asymptotic exponential decay yields a time constant $34 \pm 2$ ms, presumed to be limited by the video acquisition hardware.

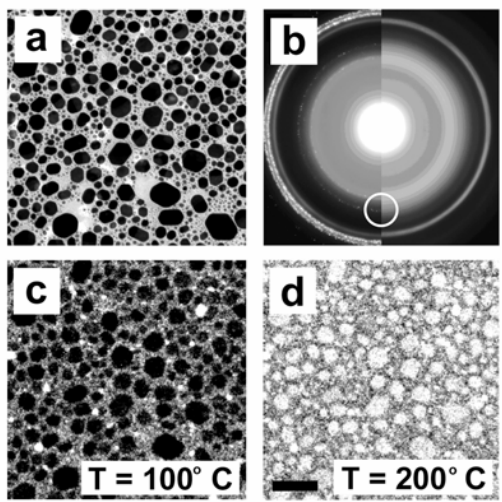

**Figure 1.**

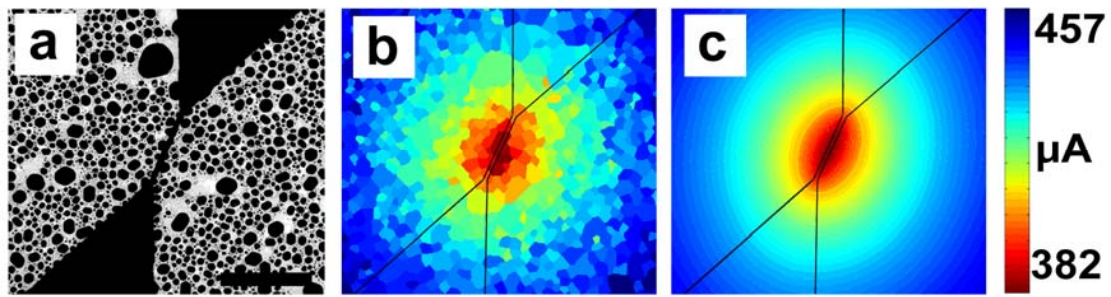

**Figure 2.**

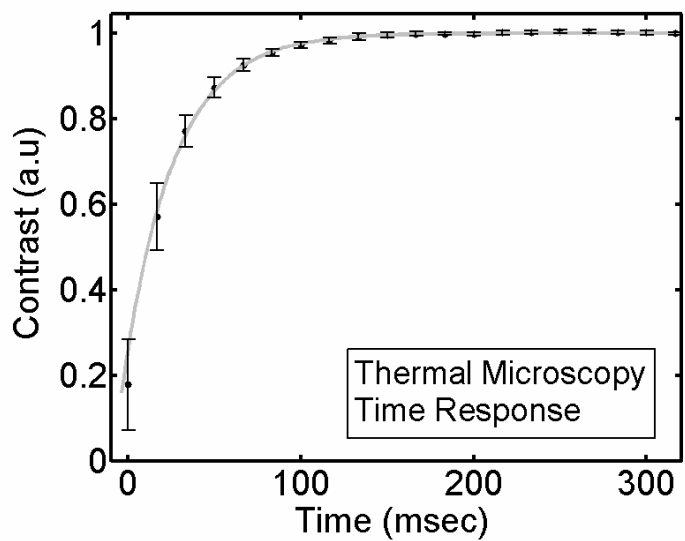

**Figure 3.**

**Supplementary Discussion for
"Electron Thermal Microscopy"**
Todd Brintlinger[1,2], Yi Qi[1], Kamal H. Baloch[3], David Goldhaber-Gordon[4], and John Cumings[1,4]

[1]Department of Materials Science and Engineering, [2]Center for Superconductivity Research, and [3]Institute for Physical Sciences and Technology, University of Maryland, College Park, MD 20742
[4]Department of Physics, Stanford University, Stanford, CA 94305

**Hysteretic behavior of islands during thermal cycling.**   In our observations we note hysteresis in the melting and freezing of the individual islands**.**  While melting of the islands proceeds uniformly, with individual islands consistently melting at the same amount of electrical power, solidification happens in a more stochastic fashion. That is to say, individual islands continue to give contrast indicative of the liquid phase despite cooling through the melting point.  This supercooling has been noted in previous studies of the melting transitions of metal islands[1-3].  To avoid possible complications of supercooling, all data shown are from the melting transition recorded upon device heating rather than the freezing transition.  Between successive increases in bias voltage, the device is allowed to cool until all islands have returned to the solid state; this occurs rather quickly as shown in the main text.

**The significance of α in the finite element model, as seen in equation (3).**  A nonzero α, also known as the temperature coefficient of resistivity or TCR, describes the transfer of potential drop—and thus Joule heating power—from the low-T, high-σ regions into the high-T, low-σ regions at increased heater current.  To render α as a fixed, dependent variable, we take advantage of the nonlinear resistance-voltage curve of our nanoscale heater wires, as shown in Figure S1.  This nonlinearity is due to the self-heating of the wire and the temperature-dependent resistivity that α characterizes. This dependence is obtained by inverting equation (3) to give $\rho = \rho_0 (1 + \alpha \Delta T)$. It is thus strongly dependent on α, and constraining the model to exhibit the experimentally observed curve safely gives $\alpha = 1.8 \pm 0.1 \times 10^{-3}$/K. This is similar to reported values[4], where the discrepancy with the bulk value[5] of $4.2 \times 10^{-3}$/K may be attributed to grain boundary or surface-scattering effects[6]. Using this value, the remaining undetermined microscopic parameters are the thermal conductivity, $K$, for the metal heater wire and for the silicon nitride support membrane. As is commonly assumed for metallic systems, the thermal conductivity of the metal wire is modeled using the Wiedemann-Franz law, leaving only the silicon nitride membrane thermal conductivity undetermined.

**Silicon nitride thermal conductivity and melting point suppression in the finite element model** As described above, the thermal conductivity of the $Si_3N_4$ membrane is left as the sole independent variable and, for simplicity, is assumed constant with temperature. The effect of the indium islands on the thermal conductivity of the $Si_3N_4$ is negligible, as we verified both by thermal modeling and by performing self-heating measurements before and after indium deposition. This thermal conductivity can then be varied in the model to find the temperature at the center of the heater wire for each of a series of applied current values. To determine the thermal conductivity, these maximum temperature values are then matched against the known heater current required to first induce melting and the known melting temperature of indium islands, $(157^{+0}_{-10}\ °C)$. Here, the uncertainty is a result of the well-characterized melting point suppression for nanoscale islands, of the size used in this study,[1,7] which we have confirmed in separate studies using a specimen heating holder with a calibrated thermocouple. The melting temperature of each island is reproducible, and in principle could be calibrated to within 1 °C or better. Instead, for simplicity, we assume that all islands have the bulk melting temperature, and treat the melting-point suppression as an experimental uncertainty. Through modeling, this translates into a $K$ value for the $Si_3N_4$ of $3.6^{+0.5}_{-0.1}$ W/m·K. Using this $K$, temperature maps at each heater current can be generated from the model, as shown in Fig. S2. The isothermal contours at the melting temperature (157 °C) are then assembled for each heater current value to give a composite thermal map, as shown in Fig. 2c. Other values of $K$ produce qualitatively different maps than those observed experimentally, providing further confirmation of its validity.

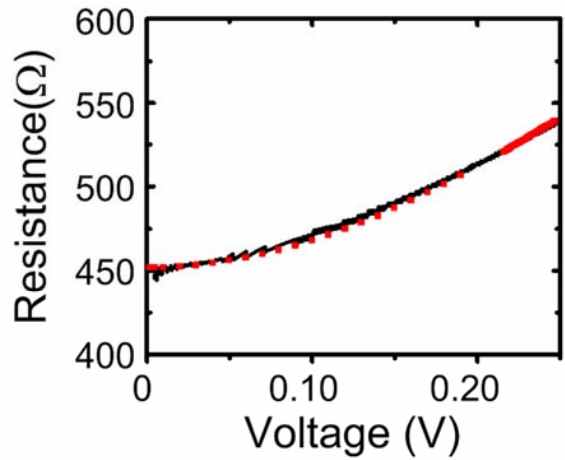

**Figure S1:** Resistance versus applied voltage curves from measurements and simulation. Larger applied voltage causes temperature to rise due to Joule heating. Resistance vs. temperature is given according to equation (3) in the main text, leaving α as a free parameter. Black line is experiment. Red squares are from simulation with α = 1.8 x $10^{-3}$/K.

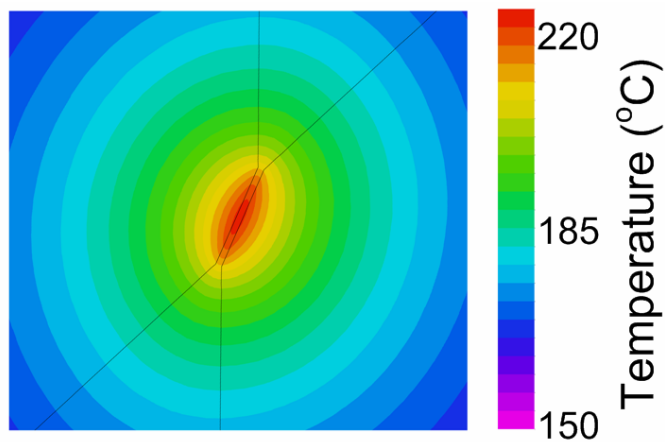

**Figure S2:** Simulated temperature distribution in Si$_3$Ni$_4$ membrane at 460μA applied current through Pd heater wire on surface. Isothermal contours are displayed with overlay of electrode location. Once *K* is obtained as described above, the temperature distribution can be calculated for each applied heater current. Figure shows same area and scale as Fig. 2 of main text.

**Supplementary Discussion References:**

1. Allen, G. L., Bayles R. A., Gile, W. W., & Jesser, W. A. Small particle melting of pure metals. *Thin Solid Films* **144**, 297 (1986).
2. Buffat, Ph. & Borel, J. P. Size effect on the melting temperature of gold particles. *Phys Rev A* **13**, 2287 (1976).
3. Zhang, M. *et al*. Size-dependent melting point depression of nanostructures: Nanocalorimetric measurements. *Phys Rev B* **62**, 10548 (2000).
4. Shivaprasad, S. M. & Angadi, M. A. Temperature coefficient of resistance of thin palladium films. *Journal of Physics D: Applied Physics* **13**, L171 (1980).
5. Smithells, C. J., Gale, W. F., & Totemeier, T. C. editors *Smithells metals reference book* (Elsevier Butterworth-Heinemann, Amsterdam ; Boston, 2004), p. 19-2.
6. Menke, E. J., Thompson, M. A., Xiang, C., Yang, L. C., & Penner, R. M. Lithographically patterned nanowire electrodeposition. *Nat Mater* **5**, 914 (2006).
7. Wronski, C. R. M.. Size dependence of melting point of small particles of tin. *British Journal of Applied Physics* **18**, 1731 (1967).